\documentclass[amsmath,amssymb,
showpacs,twocolumn, nobibnotes,aps,prb]{revtex4-1}

\usepackage{graphicx}
\usepackage{dcolumn}
\usepackage{bm}
\usepackage{enumerate}
\usepackage{amsmath,amssymb,amsfonts}
\usepackage{verbatim}
\usepackage{color}
\usepackage{mathrsfs}
\usepackage{setspace}
\usepackage{mathtools}
\usepackage{braket}
\usepackage{lipsum}

\newcommand{\be}{\begin{equation}}
\newcommand{\ee}{\end{equation}}
\def\ba{\begin{aligned}}
\def\ea{\end{aligned}}

\newcommand{\bea}{\begin{eqnarray}}
\newcommand{\eea}{\end{eqnarray}}

\newcommand{\la}{\langle}
\newcommand{\ra}{\rangle}

\newcommand{\Tr}{{\rm \, Tr\,}}

\begin{document}

\title{Structural properties of local integrals of motion across the many-body localization transition via a fast and efficient method for their construction }
\author{Safa Adami$^1$}
\author{Mohsen Amini$^1$}
\email{msn.amini@sci.ui.ac.ir}
\author{Morteza Soltani$^1$}

\affiliation{$^1$ Department of Physics, Faculty of Physics, University
of Isfahan, Isfahan 81746-73441, Iran}

\begin{abstract}
Many-body localization (MBL) is a novel prototype of ergodicity breaking due to the emergence of local integrals of motion (LIOMs) in a disordered interacting quantum system.
To better understand the role played by the existence of such macroscopically LIOMs, we explore and study some of their structural properties across the MBL transition.
We first, consider a one-dimensional XXZ spin chain in a disordered magnetic field and introduce and implement a non-perturbative, fast, and accurate method of constructing LIOMs.
In contrast to already existing methods, our scheme allows obtaining LIOMs not only in the deep MBL phase but rather, near the transition point too.
Then, we take the matrix representation of LIOM operators as an adjacency matrix of a directed graph whose elements describe the connectivity of ordered eigenbasis in the Hilbert-space.
Our cluster size analysis for this graph shows that the MBL transition coincides with a percolation transition in the Hilbert-space. By performing finite-size scaling, we compare the critical disorder 
and correlation exponent $\nu$ both in the presence and absence of interaction. Finally, we also discuss how the distribution of diagonal elements of LIOM operators in a typical cluster signals the transition.
\end{abstract}
\maketitle
\section{Introduction~\label{Sec01}}
Currently, there is a great scientific interest to gain deeper insight into the localization phenomena in many-body quantum systems.
By now, it has been found that, in one-dimension (1D), an isolated interacting system of fermions that is subject to
quenched disorder can undergo a phase transition from a thermal regime where transport is diffusive or subdiffusive \cite{Mirlin2005,BAA,Agarwal15,Anto2016,Schulz,Panda} to the MBL phase where the transport coefficients are exponentially small in the system size~\cite{Oganesyan,Znidaric,Bardarson,Gopalakrishnan} and memory of the initial state is retained to arbitrarily long times.
 
It is thought that the MBL phase of such systems can be described in terms of emergent LIOMs which form a complete set of quasilocal conserved quantities~\cite{Serbyn2013,Huse2014,ROS}. In the absence of interaction, such a system exhibits Anderson localization~\cite{Anderson} for an arbitrarily small amount of disorder. The corresponding single-particle wavefunctions are exponentially localized in real space over a characteristic length scale which is called localization length.
In this case, a complete set of LIOMs can be identified by the occupancies of these single-particle orbitals~\cite{ROS}.
Upon turning the interaction on, multiparticle resonances start to proliferate and, hence, a stronger disorder is needed to keep the system localized~\cite{Oganesyan}.
However, if the disorder strength is sufficiently larger than the interaction strength, 
the system remains in the MBL phase and LIOMs can be understood as weakly dressed single-particle orbitals~\cite{ROS,Imbrie}.

In general, the number of ways in which a set of LIOMs can be arranged is very large, and, therefore,  the calculation of all LIOMs is a complicated task practically.
It was first pointed out that a complete set of LIOMs for a finite-size system can be obtained via labeling the eigenstates of the system by their corresponding LIOM-eigenvalues uniquely~\cite{Serbyn2013, Huse2014}.
Then, it was suggested to construct LIOMs (which do not form a complete basis) by computing an infinite-time average of initially local operators~\cite{Chandran, Geraedts}.
In this regard, various approaches like using Monte-Carlo stochastic method~\cite{Stephen}, exact diagonalization techniques~\cite{Rademaker, Abanin2016,Goihl, Peng} , and tensor networks~\cite{Chandran2015, Pekker, Pollmann, Pal2017} have 
been developed. 

Although the above-mentioned construction algorithms for LIOMs have made great progress, developing and implementing
a  simple method that allows constructing a complete set of LIOMs with the following properties simultaneously is of major interest. 
A method that (i) is non-perturbative and provides quasi-local LIOMs that commute strictly with the Hamiltonian, (ii) not essentially requires strong disorder intensity, and (iii) costs much less computationally but yet has enough accuracy.
In particular, the second property makes it possible to move away from the deep MBL phase toward the transition point and study some aspects of the phase transition using LIOMs.
Thus, the main objective of the following paper is two-folded. Firstly, we present an efficient scheme for computing a complete set of LIOMs in a non-perturbative manner. And, secondly, capture certain aspects of the MBL transition by considering the 
resulting LIOM operators as an adjacency matrix of a graph that represents the connectivity of the eigenbasis in the Hilbert-space and undergoes a percolation-like transition.

In this work, we describe and develop a fast method to construct a complete set of LIOMs explicitly for the random-field XXZ spin chain that can be used to study some statistical properties of LIOMs near the MBL to ergodic transition.
We perform our algorithm via arranging an optimized set of the eigenstates of the system  in a quasi-local unitary operator which maps the physical spin operators onto effective spins operators.
Such an ordered set of the eigenbasis can be obtained by assigning an integer index number to each eigenstate which determines its order in our desired set.
We recognize this index number by locating the original basis vector of the Hilbert-space on which that eigenstate has the largest absolute amplitude among all the eigenstates of the system.
Then, in the next step, we consider the resulting LIOM operator as an adjacency  matrix of a network whose elements indicate the connectivity of the eigenbasis in the Hilbert-space.
We illustrate that this network undergoes a percolation-like transition on crossing the transition from MBL into the ergodic phase.
The percolation transition can be understood within the Hilbert-space cluster size analysis of the fragmentation of the network associated with LIOMs.
Such a classical percolation analogy for the MBL transition was previously observed either by considering the Hamiltonian as a tight-binding model in Fock-space~\cite{Chalker,Chalker2} or by retaining only resonant contributions and mapping the quantum problem to rate equations~\cite{Mierzejewski}.
However, in this paper, we underline the importance of such a transition in a network associated with LIOMs which is a key concept in MBL transition. 
We further provide an analysis of how local observables on the clusters of this network can quantitatively capture the ergodic to MBL transition.

The rest of the paper is organized as follows.
In Sec.~\ref{Sec.II}, we describe our spin-1/2 model employed, and introduce our algorithm to construct LIOM operators.
Sec.~\ref{Sec.III} contains numerical results obtained by the implementation of our algorithm. We first represent the results concerning the locality of obtained LIOM operators. 
We then use the LIOM operators and show that the ergodic to MBL transition coincides with a percolation transition in a graph of eigenvectors in the Hilbert-space whose structure is described by the matrix representation of LIOMs.
To illustrate how the transition takes place we perform cluster size analysis and apply finite-size scaling to compare the percolation threshold and correlation exponent $\nu$ in the presence and absence of interaction.
We further discuss how the distribution of local magnetization of clusters may signals the transition and finally, concluding remarks are given in Sec.~\ref{Sec.IV}  

\section{Model and approach}\label{Sec.II}

\subsection{Model Hamiltonian}
We consider a standard model of MBL which is a spin-$1/2$ chain of length $L$ in a 
random magnetic field in the $z$-direction and can be written as:  
\be\label{E1}
     H = \sum_{i=1}^{L-1}   J\left( {\sigma}^{+}_{i}{\sigma}^{-}_{i+1}+{\sigma}^{-}_{i}{\sigma}^{+}_{i+1}+\frac12\Delta {\sigma}^{z}_{i}{\sigma}^{z}_{i}\right)+h_{i}{\sigma}^{z}_{i}
\ee
where $\sigma^\pm_i = \sigma^x_i \pm i\sigma^y_i$ are the raising and lowering spin-$1/2$ operators and $\sigma_i^{x,y,z}$ denote the Pauli operators acting on spin $i$.
Here, we use open boundary conditions and fix the exchange interaction coupling at $J=1$. The values $h_i$ are, also, drawn independently  
from a random uniform distribution $[-W ,W]$ and the parameter $\Delta$ determines the anisotropy
of the model. 
This model is known to undergo a phase transition at a critical disorder strength $W=W_c= 3.5\pm 0.5$
from an ergodic phase to an MBL phase which depends on energy density~\cite{Pal-Huse,Alet2015}. 
In the current study, we focus on the MBL side of the transition, $W>W_c$, in which the existence of LIOMs prevents thermalization. 
Using the Jordan‐Wigner transformation~\cite{JW}, this model can be mapped to a model of spinless fermions and
we are interested in two different cases when $\Delta=0$ and $\Delta=1$ which corresponds to 
the non-interacting Anderson model and an interacting and disordered fermionic model respectively.

\subsection{Approach}\label{Approach_sec}
To begin with, let us review the basic idea behind the LIOM scheme.
We first consider a non-interacting system in which $\Delta=0$. 
Upon diagonalization of the Hamiltonian in Eq.(~\ref{E1}), one obtains a set of energy eigenvalues that uniquely identifies the system’s eigenstates.
In this system, which is equivalent to a single-particle Anderson model, eigenstates are exponentially localized around some localization center and their occupation numbers are mutually commuting, conserved quantities, and hence, can form a complete set of LIOMs. These are the number operators,
\be
\label{E2}
n_\alpha=\sum_{ij}\psi^*_\alpha(i)\psi_\alpha(j)c^\dag_i c_j,
\ee
in terms of which the Hamiltonian can be rewritten as
\be
\label{E3}
H=\sum_\alpha 2\varepsilon_\alpha n_\alpha - \sum_\alpha \varepsilon_\alpha,
\ee
where the last term on the right hand side is the vacuum constant energy shift. It is now straightforward to define the corresponding LIOM operators in terms of the original spin operators via the Jordan-Wigner string operator as 

\be
\label{E4}
\tau_\alpha^{z}=2\sum_{ij}\psi^*_\alpha(i)\psi_\alpha(j)\sigma^{+}_i\left( \prod_{k={\rm min}(i,j)}^{{\rm max}(i,j)}\sigma^{z}_k\right)\sigma^{-}_{j}-1,
\ee
which allows to write the Hamiltonian as
\be
\label{E5}
H=\sum_{\alpha}\varepsilon_\alpha \tau_\alpha^{z}.
\ee

Given the locality of the $\tau_\alpha$, one could as well associate an index $i$ of the lattice to each index $\alpha$, for example considering the maximum of $|\psi_\alpha(i)|^2$.

In the presence of interactions, however, the basic idea behind the LIOMs scheme is to find a unitary transformation $U$ that defines a similar complete set of independent pseudospin-$1/2$ operators
\be\label{E6}
\tau_i^z=U\sigma_i^z U^\dagger.
\ee
With the above considerations, the following properties are fulfilled by the $\tau_i^z$ operators\cite{ROS}: 

(i) $\tau_i^z$'s are quasi local operators, in the sense that:
\be
\label{E7}
||[\tau_i^z,\sigma^a_j]||<c e^{-|i-j|/\xi},
\ee
for $a=+,-,z$ and some $\xi, c$.

(ii) $\tau_i^z$ are exactly conserved: $[H,\tau_i^z]=0$.

(iii) $\tau_i^z$ have eigenvalues $\pm 1$ ($(\tau^{z})^2=1$) and each subspace has exactly dimension $2^{L-1}$.

By the locality of the terms of $H$ and of $\tau^z_i$, it follows that $J_{i_1,...,i_a}$ are local, in the sense that they decay exponentially as a function of any couple of indices. Therefore, the Hamiltonian $H$, in the MBL phase, takes the form 
\be\label{E8}
     H = \sum\limits_i {{\varepsilon _i}{{\tau} _{i}^z} + \sum\limits_{ij} {{J_{ij}}} } {{\tau} _{i}^z}{\tau _{j}^z} + \sum\limits_{ijk} {{J_{ijk}}} {{\tau} _{i}^z}{{\tau} _{j}^z}{{\tau} _{k}^z} + ...,
\ee
where the number of different sums is $L$. The unitary $U$ is a composition of local unitary transformations as described by Refs~\cite{Imbrie2,Anto}.
With the same unitary transformation one can also define $\tau^{\pm}_i$, which complete the Pauli algebra.

It is obvious, that each arbitrary arrangement of the eigenvectors of Hamiltonian $H$ in the unitary matrix $U$ of Eq.~(\ref{E6}) results in a new set of $\tau_i$ operators  
which satisfy the above-mentioned properties (ii) and (iii) by default.
However, we are interested in finding a complete set of $\tau_i$ operators that fulfills also the quasi-locality requirement which is defined in Eq.~(\ref{E7}).
Therefore, our goal is to identify a specific arrangement of the eigenstates in $U$ that best fulfill properties (i)-(iii) altogether.

For all choices of $W$ and $\Delta$, total magnetization is a conserved quantity which implies the conservation of the $z$-component of the
total spin, $[S_t^z,H]=0$ with $S_t^z=\sum_{i=1}^L S_i^z$. Therefore, it defines a good quantum number and we can consider different magnetization sectors separately. 
Throughout the paper we use the standard notation $|n\rangle\equiv|S_1^z,S_2^z,...,S_L^z\rangle$  with $S_i^z=\uparrow,\downarrow$ for the basis states in real space. In this notation, $n$ in $|n\rangle$ is a decimal integer 
can be obtained from the $L$-bit binary representation $|n_1 n_2 ... n_L\rangle$ as $n=\sum_{i=1}^{L} n_i 2^{i-1}$ where $n_i=0,1$ stands for $S_i^z=\downarrow,\uparrow$ respectively.
Using these basis states, we consider an initial set of the basis vectors $\{|n\rangle\}$ in such a way that the Hamiltonian $H$ is block diagonalized and each block corresponds to a subspace with a fixed magnetization.
Here, we are interested to introduce an efficiant way of ordering the LIOM basis states, and hence, we will use the same labeling scheme in which each basis state can be shown by  $\widetilde{|n\rangle}\equiv|\tau_1^z,\tau_2^z,...,\tau_L^z\rangle$ with effective pseudo spin $\tau_i^z=\tilde{\downarrow}, \tilde{\uparrow}$.
Again, it is convenient to obtain the corresponding integer $\tilde{n}$ for each LIOM basis vector from its binary representation as before. 
 In what follows, we introduce an optimal ordered set of basis states that makes 
the unitary operator 
\be
\label{U1}
U=\sum_{n} \widetilde{|n\rangle} \langle n |
\ee
and can be used in Eq.~(\ref{E6}) to form a complete set of LIOMs with our desired properties (i)-(iii).

We begin to construct our own approach by considering the non-interacting case, ($\Delta=0$).
In this case, as we already mentioned, LIOMs can be characterized by conserved occupations of single-particle eigenstates
and hence, only the first term on the right-hand side remains in Eq.~(\ref{E8}).
We start with the reference state $|0\rangle$ with all spins down as the only possible state in its magnetization sector which is also an eigenstate of the system. Therefore, it has a similar representation on both original and LIOM basis, i.e. 
$\underbrace{|\tilde{\downarrow} \tilde{\downarrow} ...\tilde{\downarrow} \rangle}_L = \underbrace{|\downarrow \downarrow ...\downarrow \rangle}_L$.

By flipping one spin in $|0\rangle$, we get a new state with $S_t^z=L/2-1$ and since we have $L$ places for this spin, we have $L$ states in this sector. 
These states, which are supposed to be ordered according to their binary code, form the original basis stats spanning the single-particle block of the Hamiltonian $H$. 
After diagonalizing Hamiltonian $H$, we obtain a set of eigenstates $|\psi_m\rangle$ which needs to be ordered.
According to the quasi-locality criterion of Eq.~(\ref{E7}), we expect an ordered set in which each pseudospin operator $\tau_i^z$ is mostly localized around a physical spin operator in real space.
Therefore,  we can order the obtained eigenstates by determining their maximum overlap with the original basis states. For instance, the first eigenstate is the one that has maximum overlap with the first original basis state. 
Thus, we need to find the maximum available overlap among the set
$\{|\langle \uparrow \underbrace{\downarrow \downarrow ... \downarrow}_{L-1}|\psi_m \rangle |^2,\;m=1,...,L\}$. 
If the $m_0$-th eigenstate is the one with maximum overlap with the first original basis state, it is the first basis state in psudo spin space
which means that $|\tilde{\uparrow} \underbrace{\tilde{\downarrow} \tilde{\downarrow} ... \tilde{\downarrow}}_{L-1}\rangle = |\psi_{m_0}\rangle$.
By the same token, it is possible to determine the $j$th-eigenstate of this sector by defining the following sequence of eigenstate overlaps, 
\be
\label{E10}
\{\alpha_m^j = |\langle \downarrow ... \underbrace{\uparrow}_{\text{$j$th}}   \downarrow |\psi_m \rangle|^2\}, \;\;  m=1,...,L\;,
\ee
and finding the eigenstate which maximizes the above overlap and label it with
$|\tilde{\downarrow} ... \underbrace{\tilde{\uparrow}}_{\text{$j$th}}   \tilde{\downarrow} \rangle$.

We now proceed to the next sector which has $L(L-1)/2$ basis states with two flipped spins which can be represented by the following notation,
\be
\label{E11}
|j_1,j_2\rangle = S_{j_1}^+ S_{j_2}^+ |0\rangle=|\downarrow... \underbrace{\uparrow}_{\text{$j_2$th}} ... \underbrace{\uparrow}_{\text{$j_1$th}} ... \downarrow \rangle.
\ee
The indexes $j_1$ and $j_2$ immediately determine the associated integer number $n$ of this basis vector accordingly.
Therefore, if we are looking for the $\tilde{n}$-th eigenstate in our optimized set, we should find the one with maximum overlap with its corresponding basis vector. That is finding the maximum value among the following set of overlaps
\be
\label{E12}
\{ |\langle j_1,j_2 | \psi_m \rangle|^2,\;\; m=1,...,L(L-1)/2\}, 
\ee
and label it as $\tilde{n}$.
The same analysis can be performed in higher-excitation sectors to 
arrange the final ordered set of the eigenstates properly.

Besides its ease of use and implementation, the main advantage of the above-mentioned algorithm is that it can be generalized even for the case of an interacting system ($\Delta\neq0$).
At the same time,  it provides an opportunity to consider the whole Hamiltonian in the full Hilbert-space of the system, simultaneously.
Therefore, In the remainder of this subsection, we elaborate on the implementation of the algorithm which allows ordering a generic set of energy eigenstates of the system in such a way that the resulting LIOMs satisfy our desired properties (i)-(iii).

\begin{figure}[t!]
  	\includegraphics[width=1\linewidth]{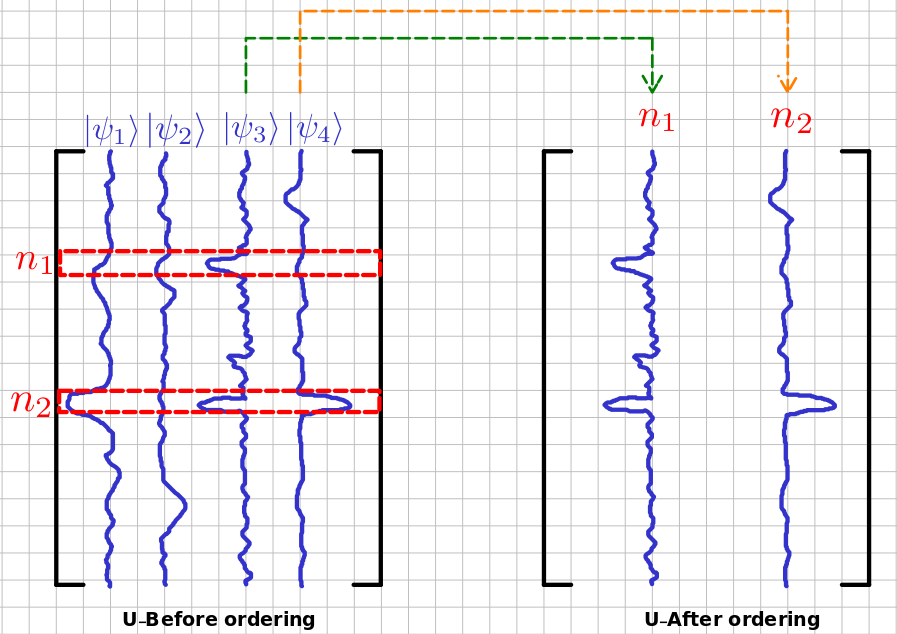}
	\caption{\label{fig-method}
	Graphical representation of the procedure outlined in section~\ref{Approach_sec} to arrange the eigenstates of the system in the unitary matrix $U$ which is obtained after diagonalization.
	The left side shows the matrix of eigenvectors obtained via the exact diagonalization procedure in which usually the eigenstates are ordered according to their corresponding eigenvalues.
	The right side is the matrix of eigenstates after rearranging them using our algorithm.
	}
\end{figure}

Suppose that we have an initial set of the basis vectors $|n\rangle$ in which our Hamiltonian matrix is block-diagonal. 
We can diagonalize this Hamiltonian and obtain the set of energy eigenbasis which is an
 arbitrary (but fixed) arrangement of energy eigenbasis of the system.
Our aim is to rearrange them by assigning a decimal integer that determines their index in our final optimum set.
Doing so, we use the fact that each eigenstate of the system is a  $2^L$ component vector which can be expanded based on the Hilbert-space original basis vectors $|n \rangle$ as $|\psi_i \rangle = \sum_{n^\prime=1}^{2^L} A_{n^\prime}^i |n^\prime\rangle$.
We can label a given eigenbasis $|\psi_{i} \rangle$ by integer number $n$ if this eigenstate has the largest absolute amplitude on $|n\rangle$ among all the eigenstates of the system. 
This means that one needs only to find the index $n$ in such a way that $|A_n^{i}|^2$ is the largest value of the set $\{|A_n^i|^2, i=1,...,2^L\}$.
In the other word, if we consider the matrix $U$ which initially contains the eigenbasis of the Hamiltonian in its columns with an arbitrary arrangement, to find the $n$-th eigenstate in our desired order, we need just to look at the $n$-th row of $U$ and determine which column has the maximum absolute value in this row. This procedure is represented graphically in Fig.~\ref{fig-method}.
The repeated execution of this procedure results in our optimal arrangement of the eigenstates which can optimally satisfy our desired conditions.
Since the procedure outlined above does not essentially require the strong disorder limit, and on the other hand, is very simple and fast we can use it
to obtain our optimum and complete set of  LIOMs rapidly and study some structrual properties of the system across the phase transition.


\section{Results}\label{Sec.III}
To examine our method we have carried out numerical calculations based on the exact diagonalization technique.
In what follows, we consider  a spin chain with $L$ spins and open boundary condition. In order to gain a deeper understanding 
of the role of interaction in the MBL case, we consider our model in both interacting ($\Delta=1$)  and non-interacting ($\Delta=0$) regimes, and depending on the system size $L$, $10^5$ to $5\times10^3$ disorder realizations are employed to obtain  
the statistics.
\subsection{Effective characterization of LIOM locality}
In this section, we demonstrate the quasi-locality of the resulting LIOM operators obtained by our algorithm.
To this end, we use the two-point correlator between a LIOM operator $\tau_j^z$ and physical spin $\sigma_k$ which is expected~\cite{Chandran,Peng,Geraedts}  to decay exponentially with distance $|k-j|$ as:
\be
\la \tau_j^z\sigma_k^z\ra= \Tr(\tau_j^z\sigma_k^z)\sim\exp{(-|k-j|/\zeta)},
\label{E13}
\ee
when $j$ and $k$ are far apart in the MBL regime.
In Eq.~(\ref{E13}) $\zeta$ defines a lengthscale over which the corresponding LIOM operators are localized. This lengthscale is related to the spatial correlation length of the eigenstate amplitudes on the Fock-space~\cite{Ivan,Logan} and expects to diverge at the critical point.

\begin{figure}[t!]
	\center{\includegraphics[width=1.1\linewidth]{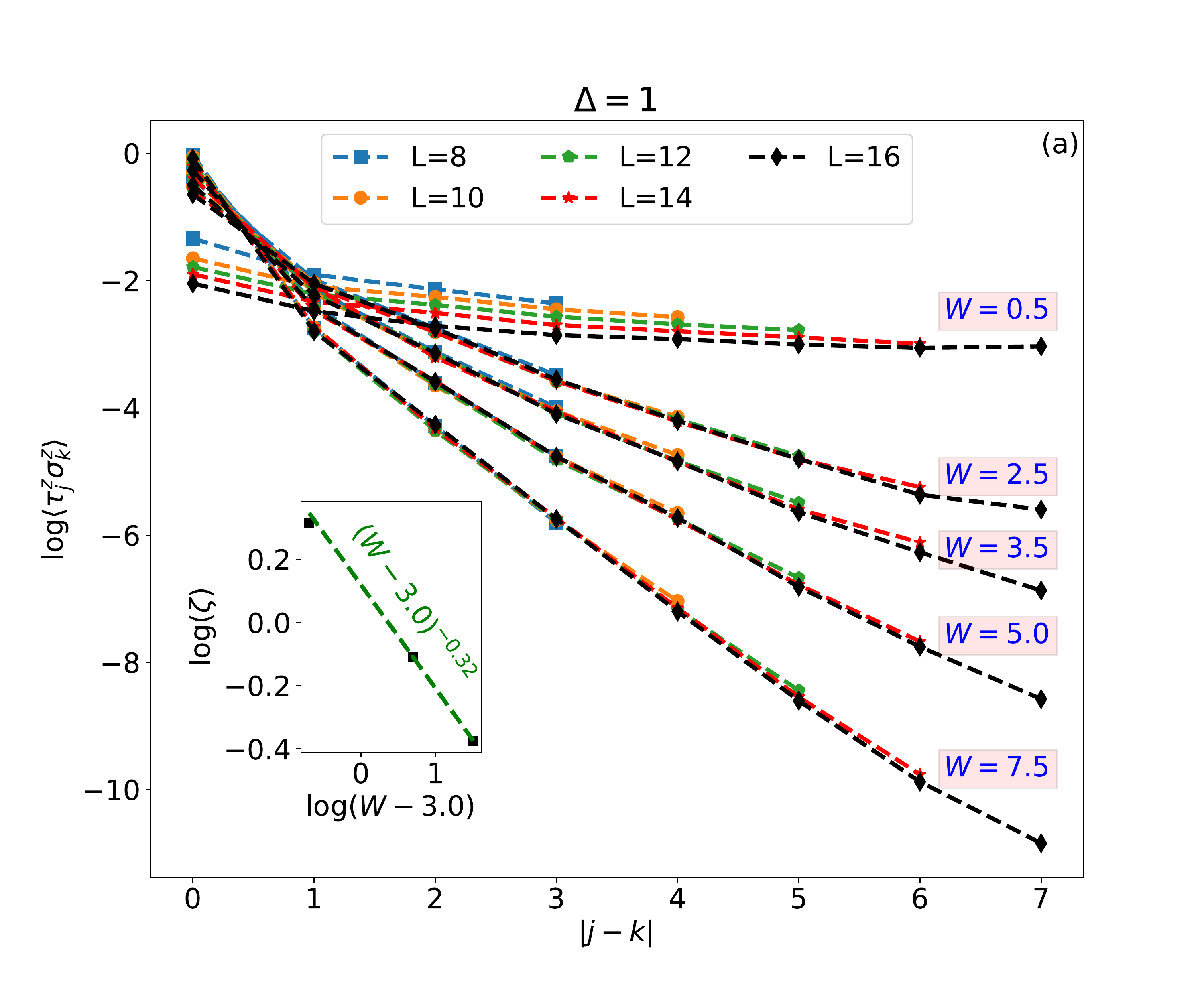}} 
         \center{\includegraphics[width=1.1\linewidth]{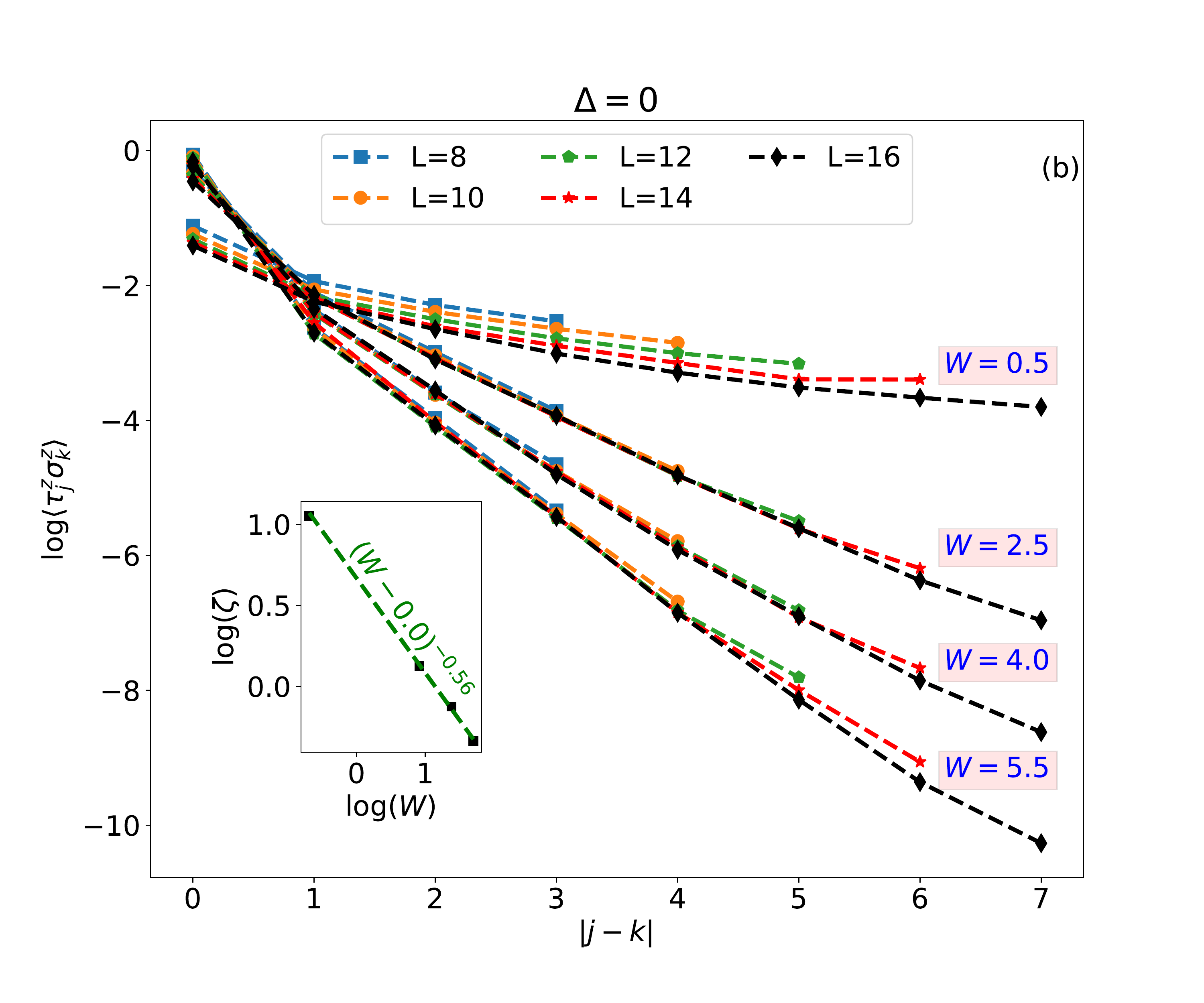}} 
        \caption{Decay of two-point correlator ($\log(\la \tau_j^z\sigma_k^z\ra)$) for the LIOM operator localized near the chain center is averaged over many disorder realizations versus $|j-k|$ for different system sizes $L=8-16$ and disorder intensities in both  (a) interacting ($\Delta=1$) and (b) non-interacting ($\Delta=0$)  regimes. Insets show the divergence of lengthscale $\zeta$ as a function of $(W-W_c)$ in the presence and absence of interaction respectively.}
	\label{fig-Localization-lengths}
\end{figure}

Fig.~\ref{fig-Localization-lengths} (a) shows the behavior of the logarithm of $\la \tau_j^z\sigma_k^z\ra$ versus $|j-k|$ both for $\Delta=0$ and $\Delta=1$.
It is obvious that in the deep localized regime $(W\ge5)$, the LIOM operators $\tau_j^z$ are strongly localized, and the $\la \tau_j^z\sigma_k^z\ra$ profile is mostly localized near the origin $j$ with a fast decaying function 
to the neighborhood.
This is in contrast to the delocalized regime $(W\le2.5)$ in which such a fast decaying  part  is obviously absent. Furthermore, there is a clear size dependency specially near the origin which is the characteristic feature of ergodic regime.
In order to make a comparison with non-interacting system $(\Delta =0)$ we have shown the behavior of the  $\la \tau_j^z\sigma_k^z\ra$ profile in Fig.~\ref{fig-Localization-lengths} (b).
In this case ($\Delta=0$), even for very small disorder strength $W=0.5$ the faster decaying behavior as well as weaker size dependency can be observed in comparison to the MBL counterpart ($\Delta=1$).

It is also worth mentioning that the characteristic lengthscale $\zeta$ can be extracted from the linear part of the $\log(\la \tau_j^z\sigma_k^z\ra)$ versus $|j-k|$ for the largest system size to observe its divergence near the transition point.
The inset shows the power-law divergence of the $\zeta$  as a function of $(W-W_c)$ on approaching the transition point ($W_c=3.0$ and $W_c=0.0$ in the presence and absence of interaction respectively) in the localized regime ($W>W_c$).

Although the above comparison between the locality of LIOMs in both interacting and non-interacting systems is a piece of qualitative evidence for their difference in the sense of critical disorder needed for localization transition, we will elaborate on this more quantitatively in the coming sections and discuss the critical disorder $W_c$ on which the transition takes place in detail.

\subsection{Percolation transition in connected clusters associated with LIOMs in the Hilber-space}
In this section, we introduce a classical percolation problem associated with clusters of LIOMs in the Hilber-space.
Indeed, LIOMs are dressed versions of spin operators as given by Eq.~(\ref{E6}) and can be viewed as a matrix in the Hilbert-space eigenbasis.
Our idea is to interpret this matrix as an adjacency matrix representation of a finite directed graph.
Under this interpretation, the nodes are the eigenbasis of the Hilbert-space and the entries in a matrix associated with each LIOM indicate whether two nodes are adjacent or not.
Obviously, to make our percolation picture we need to consider a connectivity threshold (CT) for deciding below which edge between a pair of basis states on the Hilbert-space graph will be removed which we describe how to estimate in the following.

\begin{figure}[t!]
	\center{\includegraphics[width=1.1\linewidth]{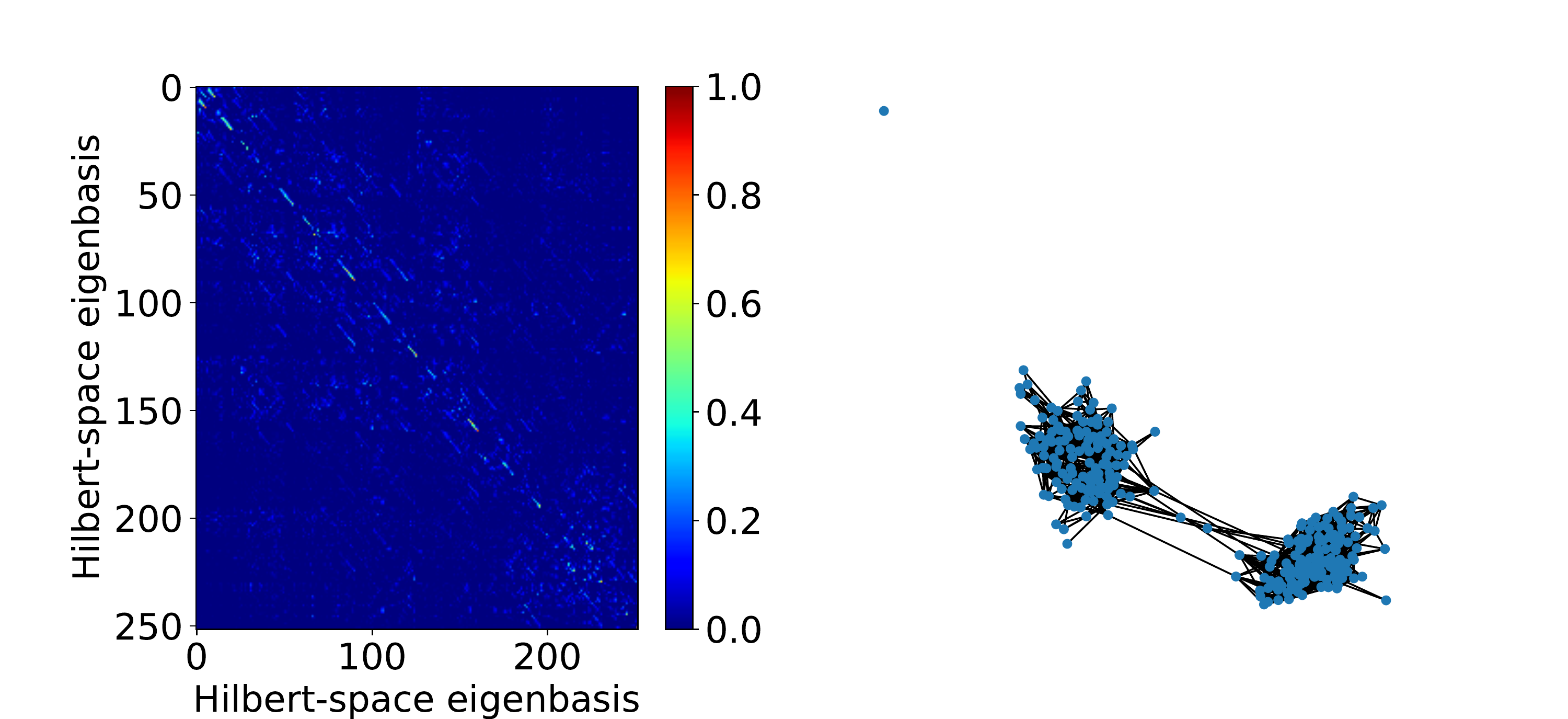}} 
	\center{\includegraphics[width=1.1\linewidth]{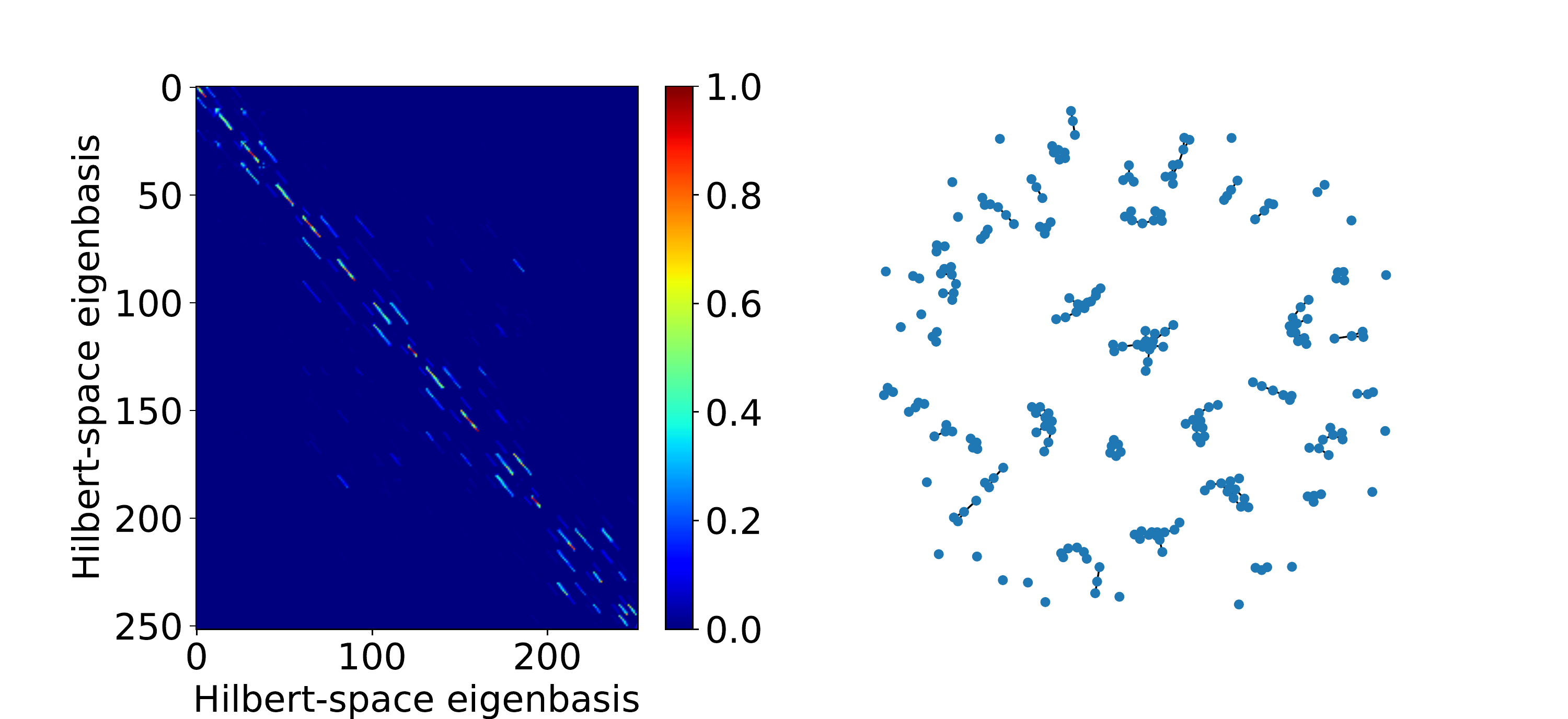}} 
	\caption{The Left side shows the matrix representation of a typical LIOM operator obtained using Eq.~(\ref{E6}) and the ordered set of eigenvectors in unitary matrix $U$. The right side is the corresponding adjacent graph of the same LIOM operator in the Hilbert-space eigenbasis  for a spin chain with length $L=10$ spins for two different disorder intensities  $W=2$ (upper panel) and $W=6$ (lower panel) in the MBL regime ($\Delta=1$).}
	\label{fig-clusters}
\end{figure}

One detail should be described before discussing the estimation procedure.
The point is to restrict our numerical calculations to the largest subspace with zero total spin, because the size of the Hilbert-space grows exponentially and total magnetization is a conserved quantity in our system. 
This subspace contains 
$N_H=\left( \begin{array}{c} L \\ \frac{L}{2} \end{array} \right)$ states (nodes). Thus, one naturally expects to have only a single connected cluster with size (number of nodes) $N_H$ for a very weak disorder intensity, namely $0<W\ll1$.
This criterion will give us an upper bound for the CT parameter. In our computation below we take the maximum possible value for CT according to its upper bound.
Fig.~\ref{fig-clusters} shows a typical LIOM operator (represented as a matrix) and its adjacent Hilbert-space graph which is obtained with $CT=0.05$ for a spin chain with $L=10$ spins  for two random realizations of disorder with disorder strengths $W=2,6$ in the MBL regime ($\Delta=1$).  

\subsubsection{Cluster size analysis}
In the theory of lattice percolation, the emergence of a spanning cluster at the percolation threshold which connects two opposite boundaries on the lattice is a measure of percolation transition~\cite{Bunde}.
It is obvious that such a definition doesn't make really sense for our considered network here.
Therefore, in this subsection, we characterize the percolation transition by analyzing the size of the largest and second-largest connected cluster~\cite{SS2} associated with LIOMs in the Hilbert-space. 
Starting from the low disorder limit, $W\ll1$, we expect to have only a single connected cluster with size $S_1=N_H$ which contains all the nodes of Hilbert-space.
Due to the localization of eigenstates, one expects to observe a decrease in the largest cluster size by increasing the disorder intensity which means that it only contains a
finite fraction of the Hilbert-space nodes. 
In the percolation language, this is equivalent to the formation of smaller size clusters in the network. 

\begin{figure}[t!]
	\center{\includegraphics[width=1.1\linewidth]{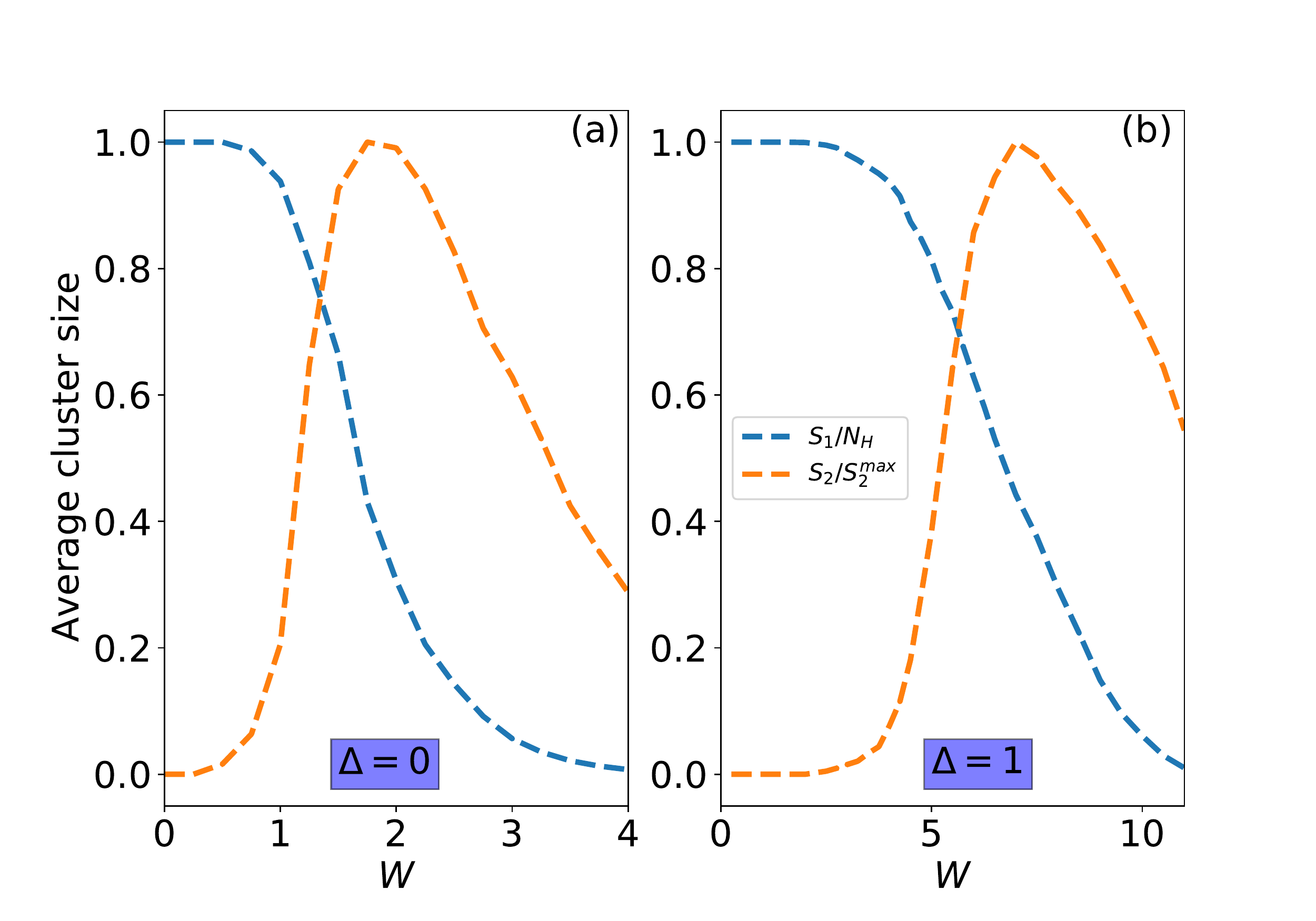} }
	\caption{Behavior of the average largest and second-largest cluster sizes versus disorder intensity $W$ for a spin chain with $L=16$ spins in both (a) the non-interacting $(\Delta=0)$ and (b) interacting $(\Delta=1)$ regimes. It is obvious one needs a stronger disorder to reach the non-percolating regime in the presence of interaction.}
	\label{S1-S2-fig}
\end{figure}

We start to illustrate our percolation scenario by calculating the fraction of the largest cluster, defined as the relative size of the largest connected cluster with respect to the Hilbert-space dimension, $S_1/N_H$.
Fig.~\ref{S1-S2-fig} shows how the mean largest cluster size which is averaged over different realizations of disorder decreases as a function of disorder intensity $W$ for a spin chain of length $L=16$.
Additionally, we can also compute the average size of the second-largest cluster $S_2$ to confirm the transition threshold.
We observe that the normalized size of the second-largest cluster, $S_2/S_2^{max}$, also peaks near the percolation transition.
We have plotted the same quantities for both non-interacting and interacting regimes in Fig.~\ref{S1-S2-fig} (a) and (b) respectively in order to make a comparison possible.
It is obvious that the presence of interaction shifts the percolation threshold $W_c$ toward the stronger intensity of disorder, however, we leave the discussion of more accurate determination of such a percolation threshold for the sections 
that follow.


\begin{figure}[t!]
	\center{\includegraphics[width=1.1\linewidth]{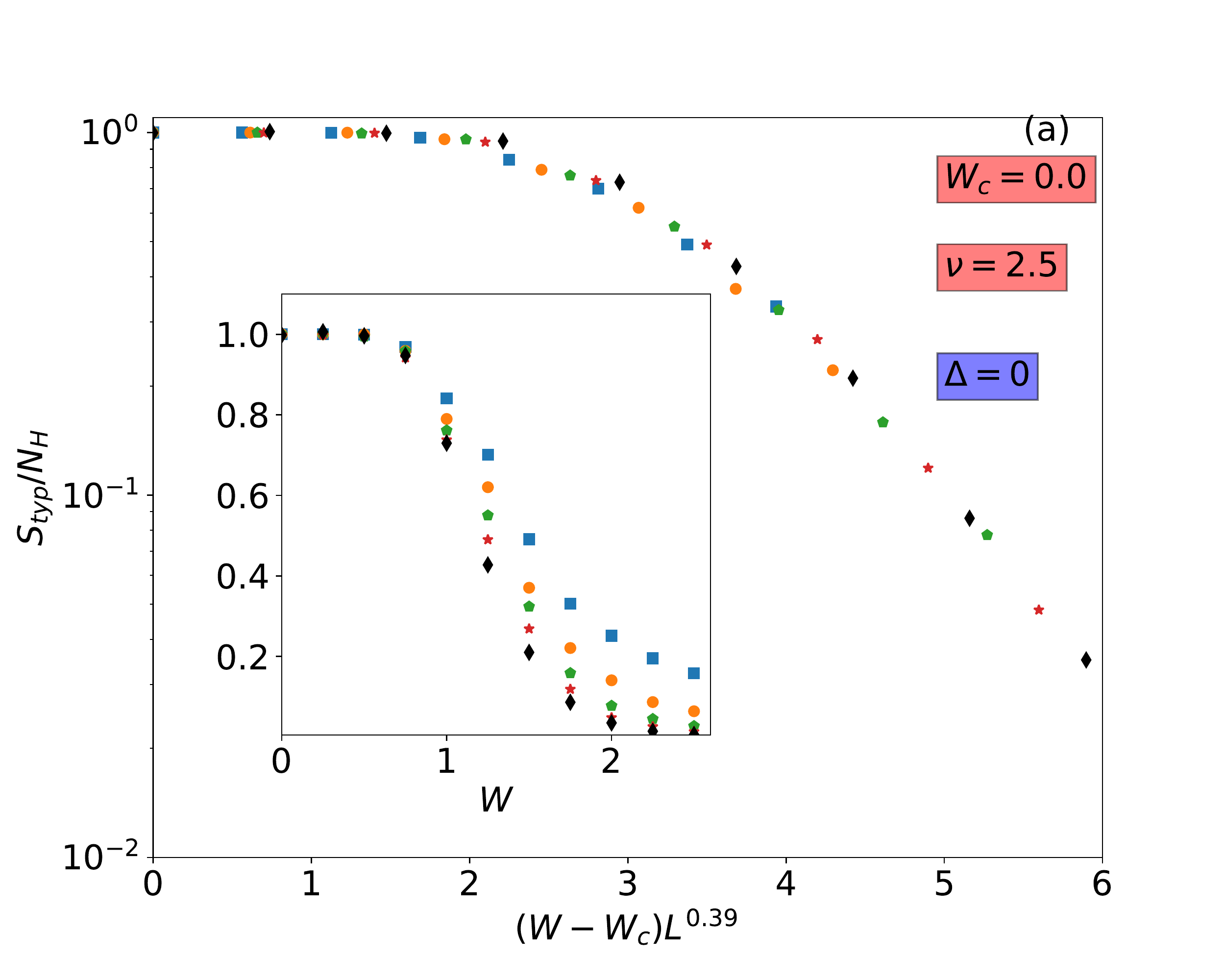}} 
	\center{\includegraphics[width=1.1\linewidth]{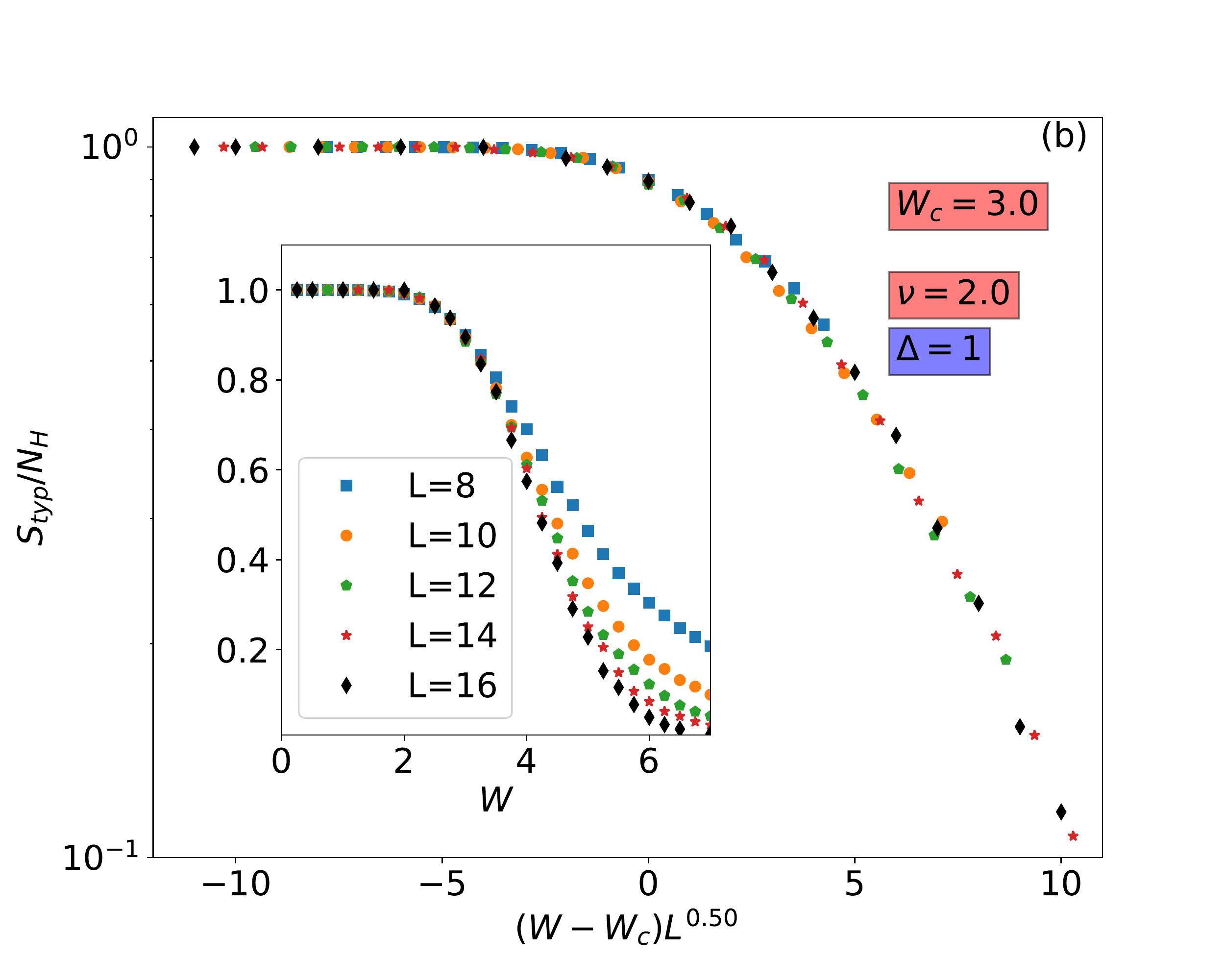}} 
	
	\caption{The resulting data collapse of $S_{typ}/N_H$  onto a scaling function of $(W-W_c) L^{1/\nu}$which is obtained for spin chains with different lengths  $(L=8-16)$ both in the (a )absence $\Delta=0$ and  (b) presence $\Delta=1$ of interaction in the localized phase. The critical parameters $W_c=3.0, 0.0$ and $\nu=2.0,2.5$ is obtained for the case of MBL and Anderson transition respectively. Insets show the corresponding raw data.}
	\label{FSS}
\end{figure}

\subsubsection{Universal feature and finite-size scaling analysis near the transition point}
The more precise determination of the percolation threshold for the adjacent graph of the LIOMs can be obtained using scaling analysis.
Following the arguments of Ref.~\cite{Chalker}, we first focus our attention on the scaling of the mean cluster size, and since the largest cluster is not essentially a typical one we take the cluster $C$ that contains the eigenstates $|\psi_0\rangle$ with the closest energy to the mean value of the energy spectrum and compute the geometric average of its size over different realizations of disorder as,
\be
S_{\text{typ}} = \exp{(\frac{1}{N_r} \sum_r \ln(s_r))}.
\ee
Here, $s_r$ is the number of nodes (eigenstates) in the cluster $C$ for a given realization $r$ of disorder and $N_r$ is the number of disorder realizations.
According to the finite size scaling~\cite{Bunde} the scaling of the normalized cluster sizes near the transition point can be stated as~\cite{Chalker}
\be
S_{\text{typ}}/N_H \sim f((W-W_c)L^{\frac{1}{\nu}}),
\ee
in which the exponent $\nu$ is called the correlation length exponent.
Therefore, by performing data collapse analysis it is possible to obtain the percolation threshold $W_c$ and critical exponent $\nu$.
The results of such data collapse, yielding the critical exponent $\nu(\Delta =1)=2.0$ and $\nu(\Delta =0)=2.5$ in the presence and absence of interaction is shown in Fig.~\ref{FSS} (a) ,(b) respectively.
We should emphasize that in the data collapse procedure in the non-interacting regime $\Delta=0$ we constrained the critical point $W_c=0.0$ as a transition point of the corresponding XX model\cite{XX}.
We note that the resulting exponent $\nu$ satisfies the Harris-CCFS bound $(\nu\ge\frac{2}{d})$~\cite{HC1,HC2} confirmed for the ergodic to MBL transition recently~\cite{Chalker,Chandran2,Anto22}.

In addition, the percolation threshold is quite different as $W_c=3.0$ and $W_c=0.0$ for the case of the interacting and non-interacting systems respectively.
It is worth mentioning, that beyond the error of our analysis, the percolation threshold for the case of the interacting system coincides with the ergodic to MBL transition point $W_c\approx 3.5$~\cite{Luitz} which shows that our system experiences a percolation transition in its corresponding LIOMs across the MBL transition.

Before ending this subsection let us shed more light on the effect of changing the estimated CT parameter in our clusterization mechanism. 
As we already discussed, we set the value of this parameter by the largest value of CT for which the size of the largest cluster is exactly equal to the Hilbert-space dimension $N_H$.
It is certainly possible to consider smaller values for CT parameter.
However, our investigations show that the changing of CT parameter may change a bit the percolation threshold $W_c$, but, the correlation length exponent $\nu$ will not change.

\subsection{Distribution of local observables on the clusters}
\begin{widetext}
\begin{figure}[t!]
	\center{\includegraphics[width=2.2\linewidth]{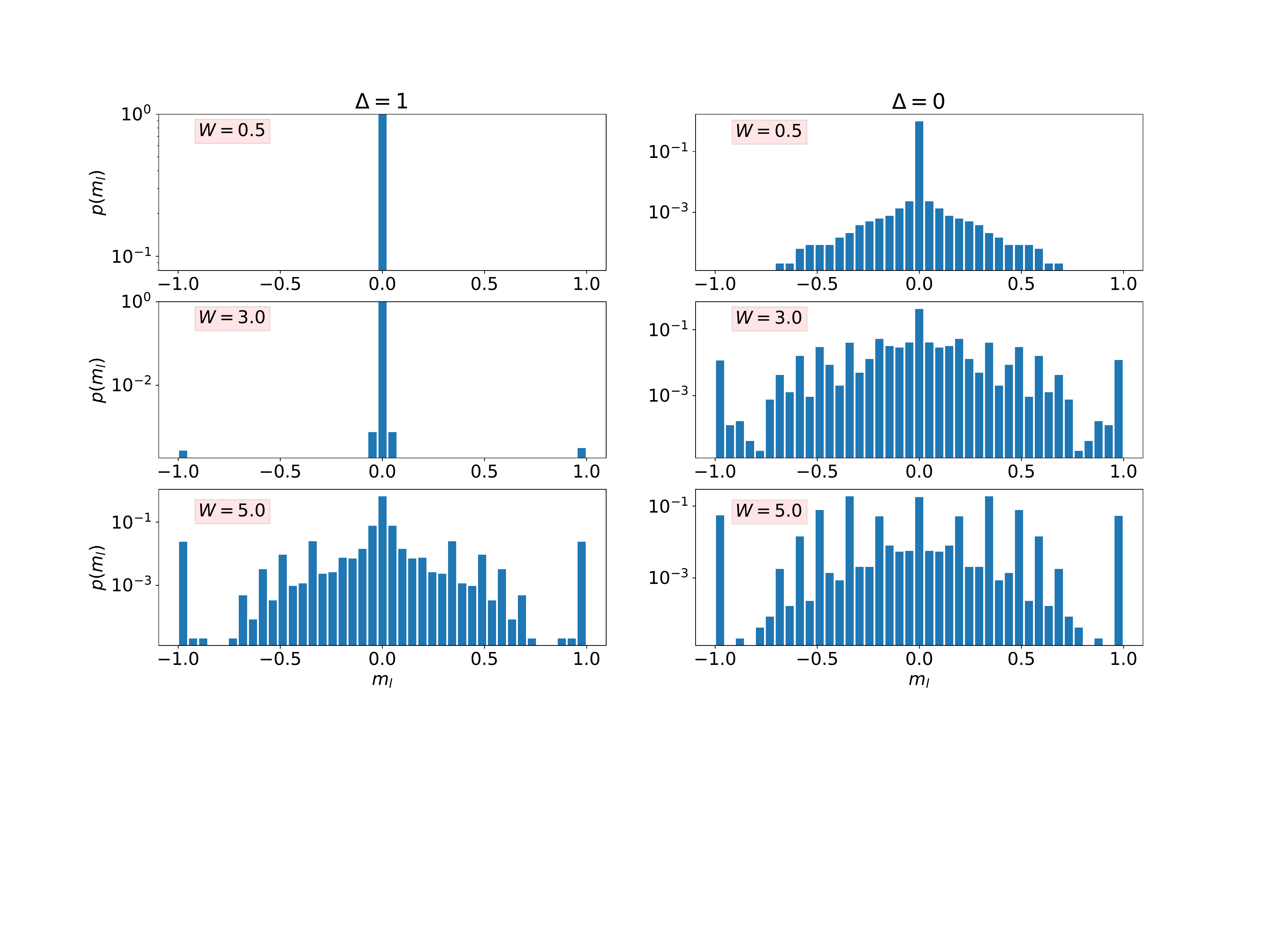}} 
	\caption{Distribution function $P(m_l)$ of the average magnetization of the cluster $m_l$ which is the average expectation value of LIOM operator $\tau^z_i$ over the nodes of this cluster both for $\Delta=0$ and $\Delta=1$.}
	\label{Diagonals}
\end{figure}
\end{widetext}

The last quantity which we are interested in is the distribution of the eigenstate expectation values of local observables which are shown to vary significantly across the MBL transition~\cite{Anto06,Luitz06}.
In doing so, let us consider the cluster $C$ which contains the eigenstate $|\psi_0\rangle$ with size $s$, as we discussed before, and define the following expectation value for the cluster~\cite{Chalker},
\be
m_l= \frac{1}{s}\sum^\prime_{\tilde{n}} \langle \tilde{n}| \tau^z_i | \tilde{n} \rangle, 
\ee
where $\sum^\prime$ denotes the restrictions imposed by considering only the eigenstates in $C$ in summation. 
This is indeed the average magnetization of the cluster and can be obtained by averaging the diagonal entries of the corresponding LIOM operator which belongs to the cluster $C$ and we are interested in the distribution of this quantity over different realizations of disorder.
Fig.~\ref{Diagonals} shows the distribution of this quantity for the largest system size $L=16$ spins in both MBL and Anderson regimes.
In the case of interacting regime $\Delta=1$ for weak disorder intensity $W=0.5$ we have a single peak around zero.
This is because in this regime the cluster $C$ contains all the eigenstates of the system which means that we are computing the trace of $\tau_i^z$ operator which we expect to be zero.
However, by increasing the disorder intensity some of the nodes (eigenstates) will be excluded from this summation, and therefore we are computing the partial trace of the local operator $\tau_i^z$. In the deep MBL phase, we expect to have clusters with smaller sizes which for the limiting case of very strong disorder one may observe clusters with only a single node with $m_l=\pm 1$. The plots in Fig.~\ref{Diagonals} show that the existence of such single-node clusters is much more probable even for very small disorder intensities in the case of the non-interacting regime ($\Delta=0$).


\section{CONCLUDING REMARKS \label{Sec.IV}}
We provided a fast and accurate method to efficiently obtain the LIOMs
for a disordered system that undergoes MBL transition. We showed that an optimized set of eigenvectors obtained by locating their maximum overlap with Hilbert-space basis can be used to obtain the desired set of LIOM operators.
We showed that the resulting LIOMs experience a percolation transition in their graph representation in the Hilbert-space by increasing disorder intensity. We also compared the critical disorder and critical exponent describing percolation transition for both interacting (MBL) and non-interacting (Anderson) regimes.  
Our analysis showed that there is a concrete connection, between the ergodic to MBL transition and the structural properties of LIOMs in their graph representation on the Hilbert-space.
\begin{acknowledgments}
We are extremely grateful to A. Scardicchio for multiple illuminated discussions, insights, and guidance to improve our work.
We acknowledge the CINECA award under the ISCRA initiative, for the availability of high performance computing resources and support.
MA acknowledges the support of the Abdus Salam (ICTP) associateship program.
\end{acknowledgments}


\end{document}